\documentclass[prl,preprint,showpacs,superscriptaddress]{revtex4}
\usepackage{amssymb,graphicx}

\begin{document}

\title{An unification of general theory of relativity with Dirac's large number
hypothesis}
\author{H. W. Peng}

\email[Email address: ] {penghw@itp.ac.cn} \affiliation{Institute
of Theoretical Physics, Chinese Academy of Sciences,
 P.O.Box 2735 Beijing 100080, China}

\begin{abstract}
Taking a hint from Dirac's large number hypothesis, we note the
existence of cosmologically combined conservation laws that work
to cosmologically long time. We thus modify Einstein's theory of
general relativity with fixed gravitation constant $G$\ to a
theory for varying $G$, with a tensor term arising naturally from
the derivatives of $G$ in place of the cosmological constant term
usually introduced \textit{ad hoc}. The modified theory, when
applied to cosmology, is consistent with Dirac's large number
hypothesis, and gives a theoretical Hubble's relation not
contradicting the observational data. For phenomena of duration
and distance short compared with that of the universe, our theory
reduces to Einstein's theory with $G$\ being constant outside the
gravitating matter, and thus also passes the crucial tests of
Einstein's theory.
\end{abstract}
 \maketitle

In Einstein's theory of general relativity as in Newton's theory
of gravitation, the strength of the gravitational interaction is
described by a
fixed dimensional constant $G_{N}\doteq 6.7\times 10^{-11}$m$^{\text{3}}$kg$%
^{-1}$s$^{-\text{2}}\doteq 7.4\times 10^{-28}$mkg$^{-1}$ $\doteq
8.2\times 10^{-45}$mJ$^{-1}$ by taking $c=3\times 10^{8}$m/s$=1$.
Einstein's theory has been applied to cosmology; another
cosmological constant with dimension of length to the power minus
two was once specially introduced by Einstein in his attempt to
construct a static model for the universe. This soon has lost his
favor as observations of the red shifts of extra-galactic nebulae
show definitely that the universe is expanding, but cosmological
constant is still used today as a free parameter in trying to fit
the observational data with the non-static homogeneous
cosmological model using the Roberston-Walker metric. Rough
estimates have thus been obtained about seventy years ago for the
age $t$\ of universe at present to be of the order of $10^{9}$
years, and recently of the order of $10^{10}$\ years. This is
about $10^{39}$ or$\ 10^{40}$ times the time $(e^{2}/m_{e}c^{3})$
needed for light to travel a distance of the classical radius of
the electron.
Similarly, with Hubble's estimate $\rho =(1.3$ to $1.6)\times 10^{-30}$ g/cm$%
^{3}$ for the density of matter due to the extra-galactic nebulae averaged
over cosmic space and a factor thousand or hundred times to include various
invisible matter, one can estimate the ratio $M/m_{p}$ of the mass $M$ of
matter in the universe of radius $R$ at present to the proton mass $m_{p}$,
and obtain another large number of the order $10^{78}$ or $10^{80}$ which is
about the square of the pervious large number. Further, the ratio $%
e^{2}/(Gm_{p}m_{e})$ of the electrostatic to the gravitational
force between the proton and the electron in an hydrogen atom is
also a large dimensionless number of the order of $10^{39}$. On
comparing these numbers, Dirac proposes \cite{Dirac1} the large
number hypothesis -- a sort of a general principle that very large
numbers which turn up in Nature and have no dimensions are related
to each other. Accepting this hypothesis and comparing the above
three large dimensionless numbers we obtain the following two
large number equations where small number coefficients are omitted
\begin{eqnarray}
&&e^{2}/(Gm_{p}m_{e})\approx
t/(e^{2}/m_{e}c^{3}),\label{eq1}\\
&& M/m_{p}\approx \{t/(e^{2}/m_{e}c^{3})\}^{2}.\label{eq2}
\end{eqnarray}
\ Dirac concludes \cite{Dirac1} from (\ref{eq1}), (\ref{eq2}) that$\ G\varpropto t^{-1}$\ and $%
M\varpropto t^{2}$ during the evolution of the universe, assuming $m_{e}$, $%
m_{p}$ and $e$\ be constant. Then neither $M\ $nor $G$\ separately but$\ $%
the combination\ $G^{2}M$ conserves in cosmologically long time.
If we take (\ref{eq1}), (\ref{eq2}) as empirically true, we have
on dividing (\ref{eq2}) by (\ref{eq1}) a relation free from
$e,m_{e},m_{p}$,
\begin{equation}
GM\approx c^{3}t\approx t,\label{eq3}
\end{equation}
on putting $c=1$. This exhibits clearly the inconsistency of applying
Einstein's theory to cosmology, because in Einstein's theory $G$\ is
constant and mass $M$ conserves, but the age of universe $t$\ does vary.

In this Letter, we take the existence of the cosmologically
combined conservation law as a hint in our attempts to generalize
Einstein's theory to a theory with varying $G$, working
consistently with the above large number relations to
cosmologically long time. In contrast to Dirac's assumption of
constant $m_{e}$, $m_{p}$ and $e$\ , we believe in the unity of
physical laws, so that $m_{e}$, $m_{p}$ will evolve in the same
way as $M$, i.e. $Gm\varpropto t$\ by (\ref{eq3}) and hence
$e^{2}\varpropto tm_{e}$ by (\ref{eq2}). Instead of Dirac's
conclusion$\ G\varpropto t^{-1}$, $M\varpropto t^{2}$ we
can only conclude that if we assume that $G\varpropto t^{-n}$, then $m$ $%
\varpropto t^{1+n}$, ($m=$\ $m_{e}$, $m_{p}$ or $M$), and
$e\varpropto t^{1+n/2}$ with $n$ likely but\ not necessarily equal
to one.\ If we make use of the constants $t_{0}$ and
$G_{0}=G(t_{0})$, we can write the proportion as an equality.
Introducing a new dimensionless variable $\phi ^{2}$ defined by
the $n$-th root of $G/G_{0}$ so that our fundamental assumption
becomes
\begin{equation}
\phi ^{2}=(G/G_{0})^{1/n}=(t/t_{0})^{-1}.\label{eq4}
\end{equation}
The above proportion laws of evolution are then converted into
cosmologically combined (c.c.) conservation laws that do remain
constant over cosmologically long time.
\begin{equation}
\widetilde{m}=\phi ^{2+2n}m=m(t_{0})\text{, }\widetilde{e}=\phi
^{2+n}e=e(t_{0}).\label{eq5}
\end{equation}%
We shall call such quantity defined by multiplication with an
appropriate power of $\phi $\ the cosmologically combined (c.c.)
quantity; it is of the same dimension as the original quantity,
and its numerical value is equal to that of the latter at the
epoch $t_{0}$, since the value of $\phi $\ at the epoch $t_{0}$\
is identically equal to one. If we choose $t_{0}$ to be the epoch
at present, then we may$\ $use $G_{N}$ for $G_{0}$.\ We shall
denote the c.c. quantity with a decoration on the letter as
exemplified in (\ref{eq5}).

We modify Einstein's theory of general relativity (here we follow
the treatment by Dirac \cite{Dirac2} with comprehensive action
principle) as follows. For matter we change the action integral
into
\begin{equation}
\widetilde{I}_{m}=-\int (g_{\mu \nu }\widetilde{p}^{\mu
}\widetilde{p}^{\nu })^{1/2}d^{4}x,\label{eq6}
\end{equation}
with the modified constraint in the form of an ordinary divergence
relation
\begin{equation}
\widetilde{p}^{\mu },_{\mu }=0,\label{eq7}
\end{equation}
where (we use the letter u for the four-velocity instead of Dirac's letter
v)
\begin{equation}
\widetilde{p}^{\mu }=u^{\mu }\widetilde{\rho }\surd
(-g),\label{eq8}
\end{equation}
with
\begin{equation}
\widetilde{\rho}=\phi ^{2n+2}\rho. \label{eq9}
\end{equation}
This constraint gives the conservation of the c.c. mass
$\widetilde{m}$ for homogeneous $\phi =\phi (t)$
\begin{equation}
\widetilde{m}=\int \widetilde{p}^{4}dx^{1}dx^{2}dx^{3}=\phi
^{2n+2}m.\label{eq10}
\end{equation}
Thus our modification of Einstein's theory consists of inserting a factor $%
\phi ^{2n+2}$ in the action integral and introducing c.c. variables. Since
Einstein's original gravitational action integral contains a denominator $%
G=G_{0}\phi ^{2n}$, so we take for the modified gravitational
action integral the expression where only a factor $\phi ^{2}$\ is
left after the cancellation
\begin{equation}
\widetilde{I}_{g}=(16\pi G_{0})^{-1}\int [\phi ^{2}R_{\mu \nu
}-w\phi _{;\mu }\phi _{;\nu }]g^{\mu \nu }\surd
(-g)d^{4}x.\label{eq11}
\end{equation}
Here we have added a kinetic term to account for the $\phi $\ to vary, the
form of the kinetic term being unique because forms like $\phi \phi _{;\mu
;\nu }$ can be transformed into $\phi _{;\mu }\phi _{;\nu }$ by an
integration by parts. The numerical coefficient $w$\ of this kinetic term
will be chosen to be $w=8$, as will be shown\ below to correspond with the
case of $k=0$ for the Robertson-Walker metric for homogeneous cosmological
models.\ We note that\ our expression (\ref{eq11}), though it looks like that
which has been used in Brans-Dicke theory of varying $G$, is actually
different from the latter because of the different relation between $\phi $\
and $G$ by our having inserted the factor\ $\phi ^{2n+2}$ in the action
integral.\ The comprehensive action principle $\delta \widetilde{I}%
_{tot}=\delta \widetilde{I}_{g}+\delta \widetilde{I}_{m}=0$ gives
the variational equations for $g_{\mu \nu }$ and $\phi $,
\begin{eqnarray}
N_{\alpha }^{\beta } &=&\phi ^{2}[R_{\alpha }^{\beta
}-(1/2)R_{\sigma }^{\sigma }\delta _{\alpha }^{\beta }]+(\phi
^{2})_{;\alpha }^{;\beta }-(\phi ^{2})_{;\sigma }^{;\sigma }\delta
_{\alpha }^{\beta }-w[\phi _{;\alpha }\phi ^{;\beta }-(1/2)\phi
_{;\sigma }\phi ^{;\sigma
}\delta _{\alpha }^{\beta }]\nonumber \\
&=&-8\pi G_{0}\widetilde{T}_{\alpha }^{\beta }=-8\pi G_{0}\widetilde{\rho }%
u_{\alpha }u^{\beta },\label{eq12}\\
\text{ \ \ \ }\Phi &=&(w\phi _{;\sigma }^{;\sigma }+R_{\sigma
}^{\sigma }\phi )=0,\label{eq13}
\end{eqnarray}
and the equations of motion $\widetilde{\rho }u^{\nu }u_{\mu ;\nu
}=0$ together with the constraint $(\widetilde{\rho }u^{\mu
})_{;\mu }=0$ for the elements of matter. These equations are not
all independent. Owing to the identities $\widetilde{T}_{\alpha
;\beta }^{\beta }=0$, which follow from the equations of motion
and the constraint for matter, and$\ N_{\alpha ;\beta }^{\beta
}+\Phi \phi _{;\alpha }=0$, which can be derived from the
invariance of the action integral $\widetilde{I}_{g}$\ by an
infinitesimal coordinate transformation as\ indicated in Dirac's
book \cite{Dirac2}, or directly verified from (\ref{eq12}) (\ref{eq13}) by
tensor calculus, these equations are compatible but indeterminate
to allow $g_{\mu \nu }$\ to change with coordinate transformation.
We note that in our theory the following
combination of (\ref{eq12}) and (\ref{eq13})%
\begin{equation}
N_{\sigma }^{\sigma }+\phi \Phi =(w/2-3)(\phi ^{2})_{;\sigma
}^{;\sigma }=-8\pi G_{0}\widetilde{\rho}\label{eq14}
\end{equation}%
is particularly simple and can be used as one of the independent equations.
Also it is interesting to write the field equations for $g_{\mu \nu }$ in a
form analogous to that in Einstein's theory by dividing both sides of (\ref{eq12})
by $\phi ^{2}$:
\begin{equation}
R_{\alpha }^{\beta }-(1/2)R_{\sigma }^{\sigma }\delta _{\alpha
}^{\beta }+\Lambda _{\alpha }^{\beta }=-8\pi G_{0}\widetilde{\rho
}u_{\alpha }u^{\beta }/\phi ^{2}=-8\pi G\rho u_{\alpha }u^{\beta
}.\label{eq15}
\end{equation}
Here on the right $G$ is not a constant but varies as $\phi ^{2n}$ while on
the left appears the tensor $\Lambda _{\alpha }^{\beta }$\ which is
completely determined by the derivatives of $\phi $%
\begin{eqnarray}
\Lambda _{\alpha }^{\beta } &=&[(\phi ^{2})_{;\alpha }^{;\beta }-(\phi
^{2})_{;\sigma }^{;\sigma }\delta _{\alpha }^{\beta }]/\phi ^{2}\nonumber \\
&&-w[\phi _{;\alpha }\phi ^{;\beta }-(1/2)\phi _{;\sigma }\phi
^{;\sigma }\delta _{\alpha }^{\beta }]/\phi ^{2}\label{eq16}
\end{eqnarray}
and differs in its tensor character from the cosmological term $%
\Lambda \delta _{\alpha }^{\beta }$ usually introduced \textit{ad
hoc} in Einstein's theory at the same place. Thus in our theory
there occurs a variable cosmological tensor but no \textit{ad hoc}
cosmological constant $\Lambda $ as a free parameter.

We now apply our theory to homogeneous cosmological model. We can adopt as
usual the Robertson-Walker metric because in simplifying the metric to this
form only considerations on symmetry and freedom of coordinate
transformation have been used, but no use is made of the field equations. We
have, as usual
\begin{equation}
\ ds^{2}=dt^{2}-R^{2}(t)\{dr^{2}/(1-kr^{2})+r^{2}d\theta
^{2}+r^{2}\sin ^{2}\theta d\varphi ^{2}\}\label{eq17}
\end{equation}%
with $t$ denoting the cosmic time and $r,\theta ,\varphi $\ the
dimensionless co-moving coordinates. For the Robertson-Walker metric the
three non-trivial field equations [one from (\ref{eq13}) and two from (\ref{eq12})
according to $\alpha =\beta =4$, and $\alpha =\beta \neq 4$]\ are connected
by one identity ($\alpha =4)$, so there are only two independent equations
to determine the two unknown function $\phi ^{2}(t)$\ and $R(t)$. We shall
choose (\ref{eq12}) with $\alpha =\beta =4$ and (\ref{eq14}) as the two independent
field equations. All the equations of motions for the element of matter are
trivially satisfied by their being at rest in the co-moving coordinates,
i.e. $u^{i}=u_{i}=0$ for $i=1,2,3$ and hence $u^{4}=u_{4}=1$. From the
modified constraint we obtain $\widetilde{\rho }_{m}=\widetilde{\rho }%
_{m}(t^{\prime })R^{3}(t^{\prime })/R^{3}$\ where the constant argument\ $%
t^{\prime }$\ is arbitrary. Then (\ref{eq14}) can be integrated to give
\begin{equation}
R^{3}d(\phi ^{2})/dt=(3-w/2)^{-1}8\pi G_{0}\widetilde{\rho
}_{m}(t^{\prime })R^{3}(t^{\prime })t,\label{eq18}
\end{equation}
the spatial factor involved in $\surd (-g)$ being cancelled out.
The constant of integration additive to $t$\ is chosen to be zero
so that $t$ is the age counted since the big bang when $R=0.$With
the help of (\ref{eq4}) $\phi ^{2}\varpropto t^{-1}$, we obtain
from (\ref{eq18}) that $R\varpropto t$, i.e. the rate $dR/dt=\dot{R}$ of
expansion is a constant. The other independent equation, namely
(\ref{eq12}) with $\alpha =\beta =4$, is for Robertson-Walker metric
explicitly
\begin{eqnarray}
&&\phi ^{2}[-3(\dot{R}^{2}+k)/R^{2}]-3(\phi ^{2})^{\cdot }%
\dot{R}/R-w\dot{\phi }^{2}/2 \nonumber\\
&=&-8\pi G_{0}\widetilde{\rho }(t^{\prime })R^{3}(t^{\prime
})/R^{3}.\label{eq19}
\end{eqnarray}
Writing proportions as equalities, for an arbitrary $t^{\prime }$
\begin{equation}
\phi ^{2}=t^{\prime }\phi ^{2}(t^{\prime })/t\text{ , \
}R/R(t^{\prime })=t/t^{\prime },\label{eq20}
\end{equation}
we see that (\ref{eq18}) is satisfied with $8\pi G_{0}\widetilde{\rho }(t^{\prime
})(t^{\prime })^{2}/\phi ^{2}(t^{\prime })=(w/2-3)$ and (\ref{eq19})\ is satisfied
with $8\pi G_{0}\widetilde{\rho }(t^{\prime })(t^{\prime })^{2}/\phi
^{2}(t^{\prime })\ =$ $w/8+3k/(dR/dt)^{2}$. These conditions agree only when
$w/8=1+k/(dR/dt)^{2}$. These conditions also show with (\ref{eq9}) and (\ref{eq4}) that $%
\rho (t)\varpropto t^{-2+n}$\ so we have $0<n<2$\ for $\rho (t)$\
to decrease with $t$\ increasing. If we believe that
$(dR/dt)^{2}\leq 1$, the choice of $w/8=1$ determining thereby
$k=0$ seems more reasonable than any other value\ chosen among the
range $w/8\geq 2$ for $k=+1$,\ while the case of $k=-1$\ is
rejected by (\ref{eq18}).

With $w=8$ and so $k=0$, the solution\ (\ref{eq20}) gives by
(\ref{eq19}) $\Lambda _{4}^{4}=2/t^{2}$, and the\ equation
(\ref{eq19}) can be expressed as $\Omega _{m}+\Omega _{\Lambda
}=1$ with $\Omega _{m}=1/3$ and $\Omega _{\Lambda }=2/3 $ in the
notation familiar with cosmologists.\ The other non-varnishing
elements of our cosmological tensor are $\Lambda _{1}^{1}=\Lambda
_{2}^{2}=\Lambda _{3}^{3}=1/t^{2}$, being one half of $\Lambda
_{4}^{4}$.

\begin{figure}[htp!]
\includegraphics[width=10cm,angle=270]{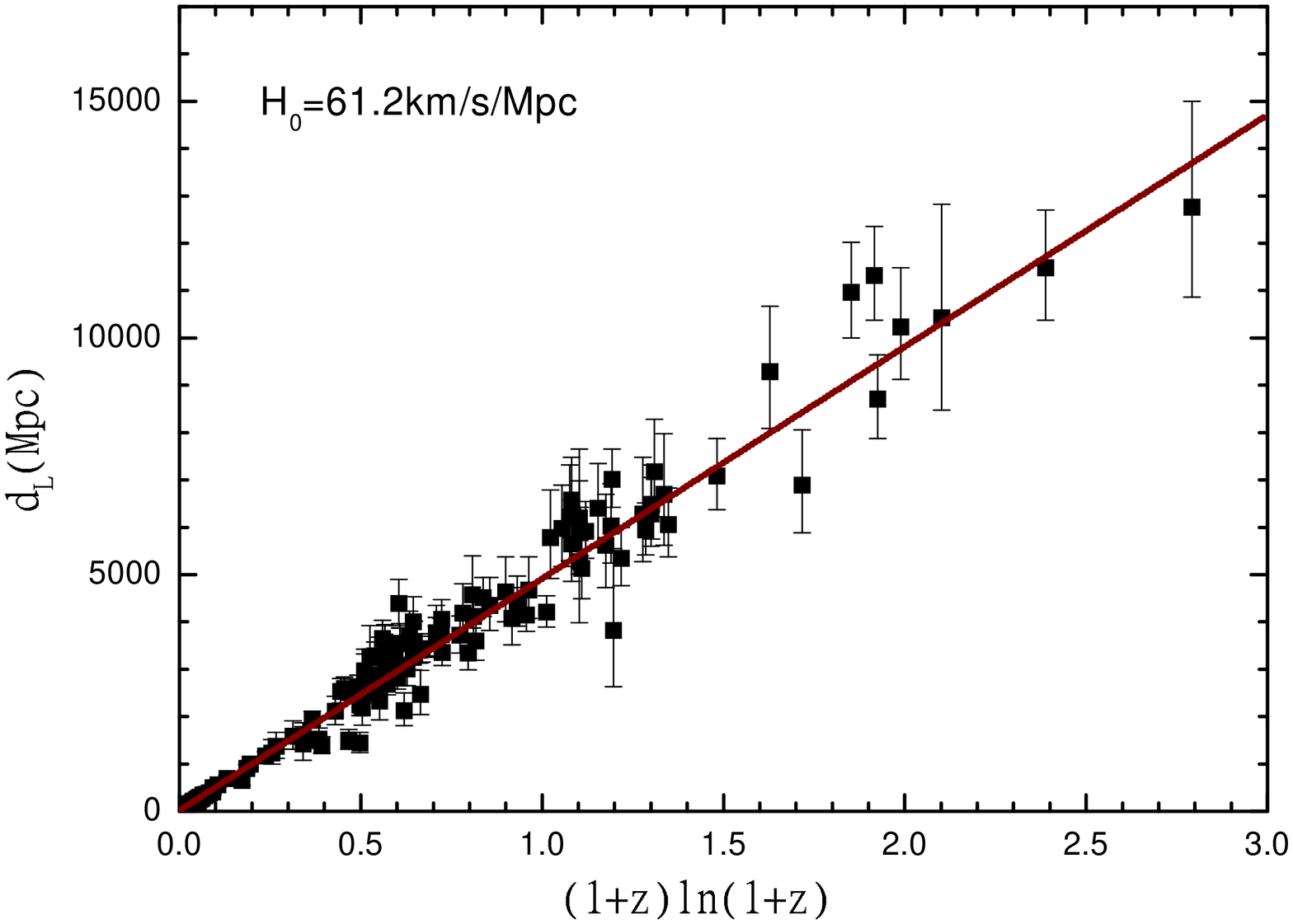}
\caption{\label{fig1} The luminosity distance and redshift. The
experimental data is taken from Ref.\cite{Riess1} and the solid
line is fitted by Eq.(\ref{eq21}). }
\end{figure}

With $k=0$ and $R\varpropto t$, we can derive the theoretical
Hubble relation very simply by following Weinberg's
\cite{Weinberg} treatment. The luminosity distance $d_{L}$ is
given by the general formula $d_{L}=(1+z)r_{1}R(t_{0})$ with
$z=R(t_{0})/R(t_{1})-1$, the suffix\ $0$ referring to the observer
at the origin $r_{0}=0$, the suffix $1$\ refereing to the source
at $r_{1}$. In
our case with\ $r_{1}R(t_{0})=R(t_{0})\int_{0}^{r_{1}}dr=R(t_{0})%
\int_{t_{1}}^{t_{0}}dt/R=t_{0}\log (t_{0}/t_{1})=t_{0}\log
[R(t_{0})/R(t_{1})]$, we obtain the theoretical Hubble relation
\begin{equation}
d_{L}=t_{0}(1+z)\log (1+z),\label{eq21}
\end{equation}
with the Hubble constant at present $H_{0}=t_{0}^{-1}$ of course.
This relation is compared with the recent experimental data of
Riess \textit{et al} \cite{Riess1} up to $z=1.55$ by Dr. Wang, as
shown in Fig.\ref{fig1}. The value of the Hubble constant $61.2$
kms$^{-1}$Mpc$^{-1}$ obtained by means of least $\chi ^{2}$ falls
within the range $H_{0}=64\pm 3$ kms$^{-1}$Mpc$^{-1}$ of their
early analysis for $z\leq 0.1$ \cite{Riess2}. The corresponding
$t_{0}=H_{0}^{-1}$ is 16 Gyr which is consistent with the
$1-\sigma$ fit by Tegmark with $\tau<0.3$ \cite{Tegmark}.

Turning to the crucial tests of Einstein's theory, which all
concern with phenomena of short duration and small distance
outside the gravitating body, e.g. our Sun, at the present epoch
$t_{0}$. We shall consider more generally
similar phenomena at the epoch $t_{e}$, say. Neglecting the difference $%
t-t_{e}$ against $t_{e}$ we can use the static or, to be precise,
the quasi-static solution of (\ref{eq12}), (\ref{eq14}) by
neglecting all time derivatives. For our purpose we need only the
exterior solution with $\widetilde{\rho }=0$ on the right hand
side of these equations. Then from (\ref{eq14}) that we see that
$\phi ^{2}$ is independent of the spatial variables, with a value
determined by the boundary condition at far from the gravitating
body at the same epoch, $\phi ^{2}=\phi ^{2}(t_{e})$. Then
(\ref{eq12}) or (\ref{eq15}) reduces to that in Einstein's theory,
as all spatial derivatives in the cosmological tensor vanishes for
constant $\phi $ and time derivatives are neglected. Thus in our
theory the static metric outside the gravitating body will be the
same
as in Einstein's theory. If we treat the sun as a mass point at rest with \ $%
\widetilde{\rho }=\widetilde{m}\delta (x)\delta (y)\delta (z)$, we
obtain the static exterior Schwarzschild solution where the
constant of integration, usually denoted by $Gm$, actually comes
from the coefficient of $\delta $ functions in
$G_{0}\widetilde{m}\delta (x)\delta (y)\delta (z)/\phi ^{2}$, so
it is the quasistatic evolving value $G(t_{e})m(t_{e})$ at the
epoch $t_{e}$, as shown by (\ref{eq4}), (\ref{eq5}). To compare
with the crucial tests of Einstein's theory, we take the present
epoch $t_{0}$\ for $t_{e}$\ and use for $Gm$\ the correspond value
at\ the present epoch, which is the same value used in Einstein's
theory. Thus we obtain the same exterior metric in our theory, as
in Einstein's theory, and quasi-static spatially constant $\phi
(t_{0})=1$\ outside the gravitating body. Hence the three crucial
tests, the advance of Mercury's perihelion, the deflection of
light by the Sun, and the gravitational red-shift of the Sun, are
all passed by our theory as by Einstein's theory, because of the
same geodesics, the same null geodesics, and the same $g_{44}$.

Finally I want to express my indebtedness to Prof. Yu Yunqiang for
drawing my attention to the luminosity distance, to Dr. Wang
Xiulian for making comparison with the data and contributing the
figure, and to Tu Zhanchun for making this manuscript to revtex4
style.

\newpage
\end{document}